\documentclass{PoS}

\newcommand{\bkp}{Broken powerlaw}
\newcommand{\lp}{Log-parabola}
\newcommand{\cop}{Cutoff powerlaw}
\newcommand{\lpcop}{\cop+\lp}

\title{The high-energy spectrum of 3C 273}

\ShortTitle{The high-energy spectrum of 3C 273}

\author{\speaker{Valentino Esposito},$^a$ Roland Walter,$^a$ Pierre Jean,$^b$ Andrea Tramacere$^a$\\
        \llap{$^a$}ISDC Data Centre for Astrophysics, ch. d'Ecogia 16, 1290 Versoix, Switzerland\\
        \llap{$^b$}Institut de Recherche en Astrophysique et Plan\'{e}tologie, 9 av. colonel Roche, BP 44 346, 31028 Toulouse Cedex 4, France\\
        E-mail: \email{Valentino.Esposito@unige.ch},
                \email{Roland.Walter@unige.ch},
                \email{Pierre.Jean@irap.omp.eu},
                \email{Andrea.Tramacere@unige.ch}}

\abstract{The high energy spectral shape of 3C 273 is usually understood in terms of Inverse-Compton emission in a relativistic leptonic jet. This model predicts variability patterns and delays which could be tested if simultaneous observations are available from the infrared to the GeV range. The instruments IBIS, SPI, JEM-X on board INTEGRAL, PCA on board RXTE and LAT on board Fermi have enough sensitivity to follow the spectral variability from the keV to the GeV and to compare them with model predictions. We are presenting preliminary results on the high energy spectrum of 3C 273 and its variability and compare these results to predictions. We found that a single component is not able to adequately fit the multiwavelength spectrum (5 keV - 10 GeV) when the statistics in Fermi data is sufficient to constraint $\gamma$-ray emission. This suggests that the X-rays emission is not originated in the jet. A possible explanation could be that X-rays are dominated by a Seyfert-like thermal inverse Compton.}

\FullConference{ An INTEGRAL view of the high-energy sky (the first 10 years) - 9th INTEGRAL Workshop and celebration of the 10th anniversary of the launch\\
                 15-19 October 2012\\
                 Bibliotheque Nationale de France, Paris, France}

\begin{document}

\section{Introduction}

\begin{figure}
\includegraphics[width=1.\textwidth]{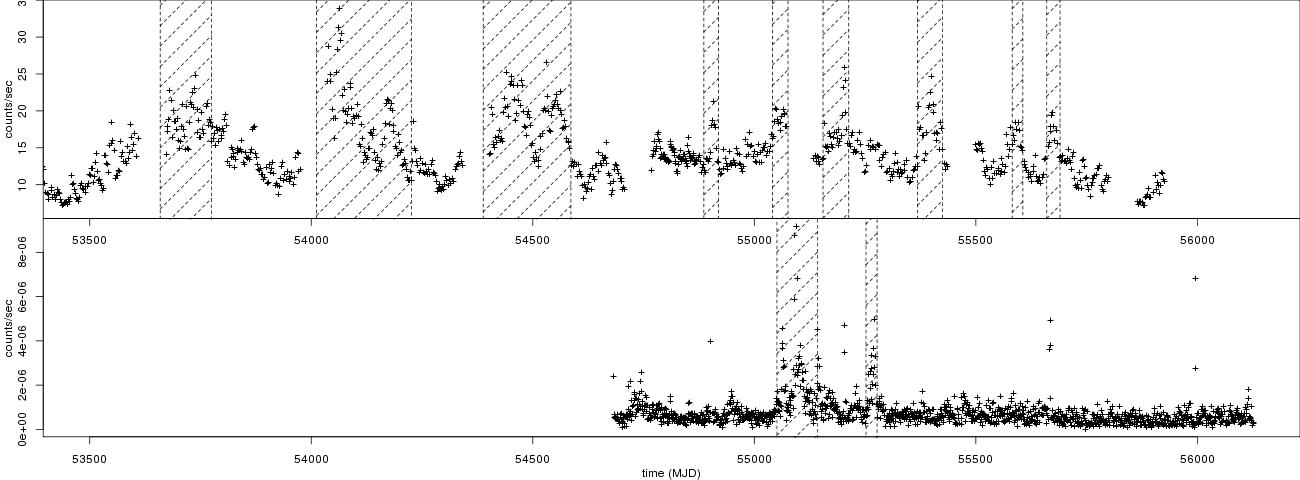}
\caption{\textit{Top panel:} PCA lightcurve in 0.2 - 75 keV. \textit{Bottom panel:} Fermi lightcurve in 0.1 - 100 GeV. Area marked by dashed lines shows the selected flaring time epochs in both lightcurves.}
\label{fltime}
\end{figure}

The bright quasar 3C 273 is one of the best monitored Active Galactic Nuclei (AGN). It shows several interesting blazar-like features (like the presence of a jet with superluminal motion and high variability) and Seyfert-like features (like a blue bump and variable emission lines). Many works have been published to describe in details its spectral and temporal characteristics, and a wide database of multiwavelength observations has been collected and it is available online.\footnote{http://www.isdc.unige.ch/3c273/} The aim of this work is to look for evidence of spectral variability related to different states of the brightness of the source. We analysed the high energy spectrum, in the keV to GeV energy range, of 3C 273 using data from RXTE-PCA (2 - 70 keV), JEM-X, IBIS/ISGRI and SPI on board INTEGRAL (3 keV - 1 MeV) and Fermi-LAT (100 MeV - 10 GeV). We selected two different epochs based on the behaviour of PCA and LAT lightcurves: the first epoch corresponds to the sum of the several flaring times in PCA while the second to the big flares in the LAT lightcurve (Fig. \ref{fltime}). INTEGRAL data were accumulated on the PCA and LAT flaring epochs to build the multiwavelength spectra of the source.

\section{Data Sample and Selection}

PCA spectra and lightcurves have been taken from the HEAVENS web interface.\footnote{http://www.isdc.unige.ch/heavens/} The signal-to-noise ratio of PCA is good enough to detect a well structured lightcurves, with clear flaring structures (Fig. \ref{fltime} \emph{top panel}). We selected the time intervals in which the source is flaring, defining the \emph{PCA flaring epoch}, i.e. when the source is flaring in X-rays.

Similarly we selected the \emph{LAT flaring epoch}, i.e. when the source is flaring in $\gamma$-rays, looking at the Fermi lightcurve (Fig. \ref{fltime} \emph{bottom panel}). The LAT lightcurve is indeed different from the PCA lightcurve: it shows a single big flare with some structures followed by another shorter flare. The first flare is described in detail in \cite{abdo2009}. Except for this outburst the count rate is relatively stable showing few structures and few isolated points with high count rate.

The INTEGRAL lightcurves do not exhibit structures like the PCA lightcurves, so they are not used to select intervals of interest. INTEGRAL data have been used for spectroscopical purpose only.

For both the two epochs described above a multiwavelength spectrum with data from all the instruments (PCA, JEM-X, IBIS/ISGRI, SPI and LAT) has been built.

\section{Spectral Analysis}

We fitted the data with several models and compared the results. We used \emph{\bkp}, \emph{\cop}, \emph{\lp} and \emph{\lpcop} model, where the \lp\ spectral law is

$$
N(E) = N_{lp} {\left( \frac{E}{1 \rm \, keV} \right)}^{-\alpha - \beta \log \left( \frac{E}{1 \rm \, keV} \right) }
$$

\noindent being $N_{lp}$ the photon flux at energy $E = 1 \rm \, keV$.

\begin{figure}
\includegraphics[width=.5\textwidth]{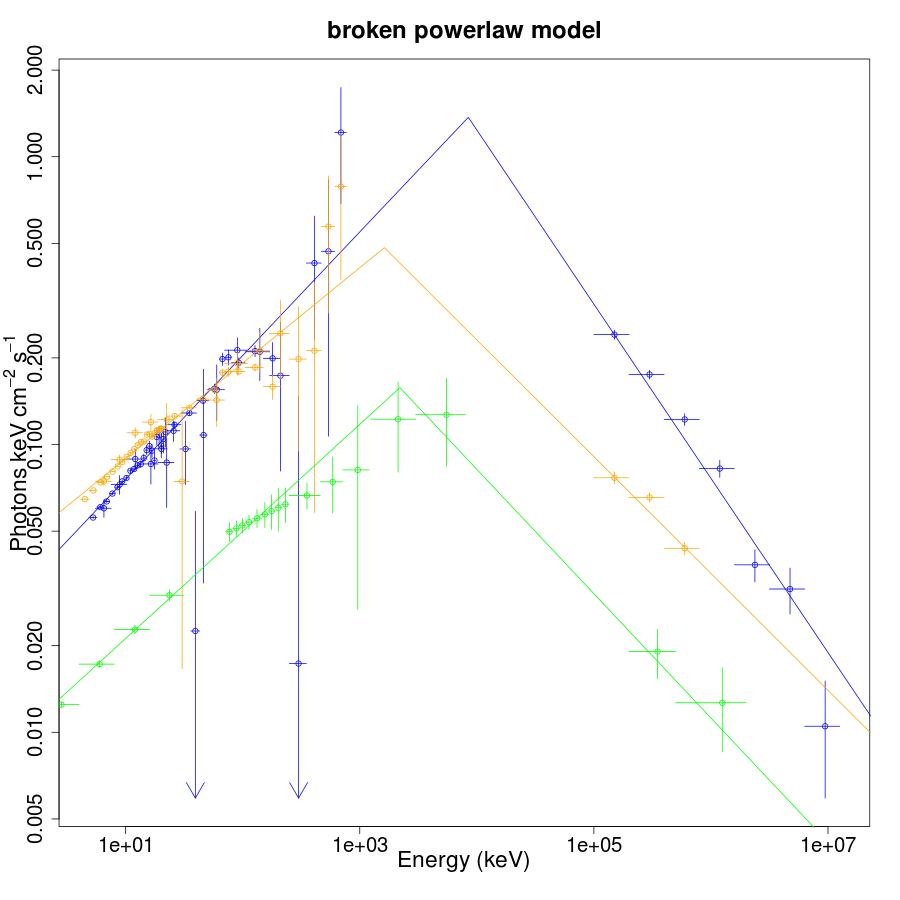}
\includegraphics[width=.5\textwidth]{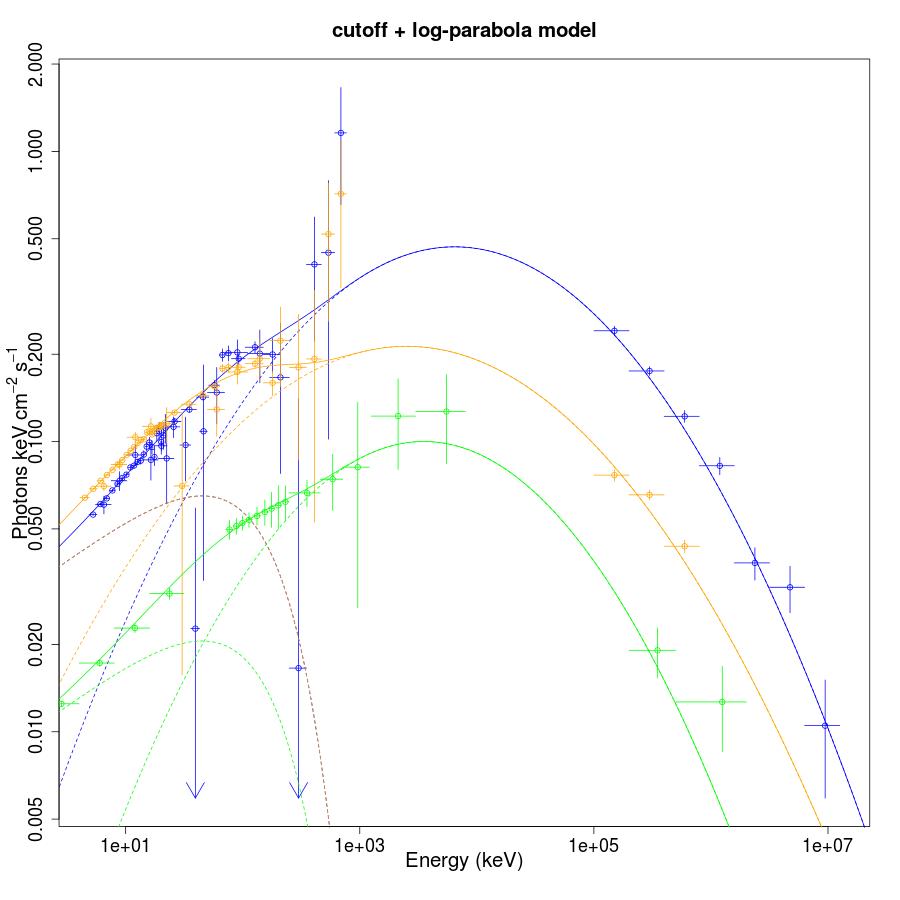}
\caption{3C 273 multiwavelength spectra fitted with \bkp\ model (\textit{left panel}) and \lpcop\ model (\textit{right panel}). Solid lines are the best fit models. In the right panel the individual components \cop\ and \lp\ are plotted with dashed lines. Blue: spectrum A; Yellow: spectrum B; Green: spectrum C. See text for details.}
\label{spec_mod}
\end{figure}

The multiwavelength spectral energy distribution (SED) with the best fit models for \bkp\ and \lpcop\ are shown in Fig. \ref{spec_mod}. In both panels, blue points and line denote the spectrum at \textit{LAT flaring time} (henceforth spectrum A), yellow points and line at \textit{PCA flaring time} (henceforth spectrum B) while the green points data taken from \emph{Litchi et al. 1995} \cite{litchi1995} are related to a multiwavelength campaign on 3C 273 performed in 1991, and they are used here for comparison (henceforth spectrum C). The observations in the X and $\gamma$-ray bands were performed by GINGA and OSSE, COMPTEL and EGRET on board CGRO. Coloured solid lines are the respective best fit models and dashed lines are the individual additive components of the model (see section \ref{sec_lpcop}). A summary of the best fit parameters for the models is reported in Table \ref{tab_par}.

\cop\ model is not discussed in detail because it cleary does not fit the data. Fermi data can not be modeled by an exponential cutoff. For the \cop\ model we obtain $\chi_{red}^2 \simeq 48$ for spectrum A and $\chi_{red}^2 \simeq 37$ for spectrum B.


\subsection{\bkp\ model}

The fit by means of the \bkp\ model does not provide a good description of the data, but still the variation of the indexes $\Gamma_1$ and $\Gamma_2$ between the states A and B can give some some clue on the physics. The indexes $\Gamma_1$ and $\Gamma_2$ are clearly different between spectra A and B, leading to a variation of the spectral break $\Delta \Gamma = \Gamma_2 - \Gamma_1$ from 0.8 (spectrum B) to 1.1 (spectrum A), and the break energy $E_b$ seems to be larger in state A than in state B, even if it cannot be well constrained because it lies in the gap between the data (between SPI and LAT). Spectrum C instead looks quite similar to spectrum B, except for the normalization which is lower.

The indexes variation between the spectra suggests a modification of the electron distribution in the jet during the $\gamma$-ray flare.

\subsection{\lp\ and \lpcop\ models}
\label{sec_lpcop}

The \lp\ is the common model used to fit blazar spectra in $\gamma$-ray energies \cite{nolan2012}. The variation of the spectral slope of the \lp\ model can represent much better the Fermi observations obtained for the spectrum A, when compared to a powerlaw with a constant slope.

The bump in the high energy part of the spectrum is commonly considered to be due to Inverse Compton (IC) processes, could be Synchrotron Self Compton (SSC) or External Compton (EC), on the high energy electrons in the jet and the \lp\ shape is able to model it. The shape may be given by a particle acceleration due to a stochastic component in the acceleration process \cite{tramacere2011}.

The \lp\ model does not provide a good description of spectrum A ($\chi_{red}^2 > 2$), because the intrinsic curvature of Fermi data in spectrum A does not match with the X-ray data in a \lp\ shape. Assuming that the \lp\ is the model for the IC bump this suggests that there is an extra component in the spectrum.

The presence of two components can be also inferred by looking at the lightcurves in Fig. \ref{fltime}. The variability pattern of the two lightcurves at different energy bands is very different, suggesting that X and $\gamma$-rays are not correlated and therefore they should be two distinct components or at least a single component with some variable parameter. In \emph{Soldi et al. 2008} \cite{soldi2008} 3C 273 variability is discussed: they suggest two distinct physical components or a single component with varying spectral shape in the X-ray emission below and above $\sim 20 \rm \, keV$ based on the lightcurve variability.

We model the extra component in the spectrum as a \cop\ at lower energies, for which we fixed the parameters in the fit. The adding of the \cop\ allows us to improve the quality of the fit. The chosen fixed values of the \cop\ parameters are indeed quite arbitrary, because if we try to fit the whole set of 6 parameters of the \lpcop\ model there is some degeneration and it is not possible to find well constrained values, especially for the cutoff energy $E_c$ which can span up to $\sim 500 \rm \, keV$ and still give a good fit.

The F-test between the two models, \cop\ and \lpcop\ on dataset A suggests that the second model is significantly better: the probability that \lpcop\ model is better by chance only is $P_f = 3 \cdot 10^{-7}$.

The spectra B and C instead can be well fitted with a \lp, because for these datasets the statistics at $E \gtrsim 1 \rm \, GeV$ (Fermi data in state B and EGRET data in state C) is poor, and the whole dataset is needed to estimate the curvature parameter $\beta$. In this case is not possible to state if there is an extra component or not from the spectrum as in spectrum A. Using the \lpcop\ model with the same fixed parameters of spectrum A the fit in spectrum B and C is still good.

The peak energy of the \lp\ SED is given by $E_{peak} = 10^{(2 - \alpha)/2 \beta} \rm \, keV$ \cite{massaro2004}. The peak energy of the three spectra seems to show a trend similar to that of $E_b$ in the \bkp\ model, with a larger $E_{peak}$ in spectrum A, and comparable values in spectra B and C. In any case the error on these parameters is quite broad so this trend may not be significant.

We also tried to fit the three datasets A, B and C togheter with the \lpcop\ and the \lp\ models, forcing the spectral parameters to be the same but leaving the normalization free for intercalibration. The different datasets can not be modeled by the same shape, proving that the difference between the datasets can not be due to the normalization only. The global fit of the two datasets A and B togheter has a $\chi_{red}^2 \simeq 13$ using both the \lpcop\ and the \lp\ model.

\begin{table}
\begin{tabular}{|c|c|c|c|c|c|c|}
\hline
& \multicolumn{6}{|c|}{\textbf{\bkp}} \\
\hline
Spectrum & $\Gamma_1$ & $E_b (MeV)$ & $\Gamma_2$ & $N_{bp}$ & $\chi_{red}^2$ & $d.o.f.$ \\
\hline
A & $1.57 \pm 0.13$ & $8.5_{-1.3}^{+1.5}$ & $2.61 \pm 0.04$ & $(2.82 \pm 0.1) 10^{-2}$ & 1.442 & 51 \\
\hline
B & $1.67 \pm 0.01$ & $1.6_{-0.8}^{+1.1}$ & $2.41 \pm 0.85$ & $(4.17 \pm 0.04) 10^{-2}$ & 2.176 & 43 \\
\hline
C & $1.62 \pm 0.01$ & $2.2_{-0.9}^{+2.9}$ & $2.43_{-0.15}^{+0.17}$ & $(0.91 \pm 0.03) 10^{-2}$ & 0.718  & 16 \\
\hline
\end{tabular}

\begin{tabular}{|c|c|c|c|c|c|}
\hline
& \multicolumn{5}{|c|}{\textbf{\lp}} \\
\hline
Spectrum & $\alpha$ & $\beta$ & $N_{lp}$ & $\chi_{red}^2$ & $d.o.f.$ \\
\hline
A & $1.22 \pm 0.01$ & $0.108 \pm 0.002$ & $(1.54 \pm 0.04) 10^{-2}$ & 2.233 & 52 \\
\hline
B & $1.41 \pm 0.02$ & $(9.74_{-0.33}^{+0.35})10^{-2}$ & $(2.91 \pm 0.19) 10^{-2}$ & 0.965 & 41 \\
\hline
C & $1.38 \pm 0.03$ & $(9.81_{-0.59}^{+0.66})10^{-2}$ & $(0.70 \pm 0.03) 10^{-2}$ & 1.290 & 17 \\
\hline
\end{tabular}

\begin{tabular}{|c|c|c|c|c|c|c|c|c|}
\hline
& \multicolumn{8}{|c|}{\textbf{\lpcop}} \\
\hline
Spectrum & $\Gamma$ & $E_c (keV)$ & $N_{cp}$ & $\alpha$ & $\beta$ & $N_{lp}$ & $\chi_{red}^2$ & $d.o.f.$ \\
\hline
A & 1.7 & 150 & 0.028 & $0.76_{-0.08}^{+0.06}$ & $0.16 \pm 0.01$ & $(2.0 \pm 0.4) 10^{-3}$ & 1.151 & 49 \\
\hline
B & 1.7 & 150 & 0.028 & $1.10_{-0.07}^{+0.05}$ & $0.13_{-0.007}^{+0.01}$ & $(6.3 \pm 1.0) 10^{-3}$ & 1.132 & 41 \\
\hline
C & 1.7 & 150 & 0.009 & $0.61_{-0.21}^{+0.20}$ & $0.19 \pm 0.03$ & $(0.34_{-0.18}^{+0.38}) 10^{-3}$ & 0.098 & 16 \\
\hline
\end{tabular}
\caption{Table of fit parameters. The first column identifies the spectrum as explained in the text. Parameters, reduced $\chi^2$ and degree of freedom are reported for each spectrum and each model. $N_{bp}$, $N_{lp}$, $N_{cp}$ are the normalizations of the respective models, corresponding to the photon flux at $1 \rm \, keV$. Errors are at the 90\% confidence level ($2.71 \: \sigma$) and the parameters without error are used as fixed parameters in the fit.}
\label{tab_par}
\end{table}

\section{Conclusion}

A careful analysis is still in progress and more models can be tested. With respect to previous work like \cite{litchi1995} or \cite{soldi2008} we have data in the $\gamma$-ray domain up to $\sim 10 \rm \, GeV$ from the Fermi-LAT spacecraft, allowing us to study the multiwavelength emission at higher energies with more detail.

A possible and interesting scenario to interpret the data is the two components scenario, in which the $\gamma$-ray emission modelled with the \lp\ spectral law could be interpreted as a blazar-like emission due to IC processes (SSC or EC) in the jet, as suggested by the correlations between the hard X-ray and radio or optical emission claimed in \cite{soldi2008}.

The X-ray component we modelled with the \cop\ could be a Seyfert-like component, due to thermal photon emitted by the accretion disk and scattered in the X-ray domain by IC scattering on a different electron population, possibly an hot corona close to the central black hole.

More detailed analysis on the lightcurves and on optical and infrared data are needed to confirm this scenario.

\end{document}